\documentclass[iop]{emulateapj}
\usepackage{amsmath}
\usepackage{amssymb}
\usepackage{graphicx}
\usepackage{epstopdf}
\usepackage{natbib}
\usepackage{color}

\begin{document}

\newcommand{\etal}{et~al.}

\title{Measuring the jet power of flat spectrum radio quasars}
\author{S. S. Shabala, J. S. Santoso}
\affil{School of Mathematics and Physics, Private Bag 37, University of Tasmania, Hobart, TAS 7001, Australia}
\author{L. E. H. Godfrey}
\affil{International Centre for Radio Astronomy Research, Curtin University, GPO Box U1987, Perth, Western Australia 6845, Australia}

\begin{abstract}

We use frequency-dependent position shifts of flat spectrum radio cores to estimate the kinetic power of AGN jets. We find a correlation between the derived jet powers and AGN narrow-line luminosity, consistent with the well-known relation for radio galaxies and steep spectrum quasars. This technique can be applied to intrinsically weak jets even at high redshift.

\end{abstract}

\keywords{galaxies: active --- galaxies: jets --- quasars: general}

\section{Introduction}

Active Galactic Nuclei (AGN) are some of the brightest objects in the Universe. This makes them visible to huge distances, and therefore useful in applications ranging from studies of Galactic ionized gas (through interstellar scintillation; \citet{LovellEA08}), to measuring positions on Earth to centimetre precision (via the technique of geodetic Very Long Baseline Interferometry; \citet{MaEA09}), to cosmological applications (through their use as standard candles; \citet{WatsonEA11}).

AGN play a key role in galaxy formation and evolution. The evolution of black holes and galaxies are tightly coupled throughout cosmic history \citep{MagorrianEA98,HasingerEA05}, and nuclear activity of black holes imparts significant feedback on the surrounding gas through radiatively and mechanically driven outflows. Radio-emitting jets of relativistic particles inflate cocoons of radio plasma, which in turn drive bow shocks through the host galaxy and beyond \citep{KA97}. On large scales, these shocks are observed to uplift the hot gas present in galaxy clusters. Even once jet activity ceases, jet--inflated radio lobes can rise buoyantly through cluster gas, transporting the gas outward \citep[e.g.][]{FabianEA03,FormanEA05}.

AGN activity is intermittent, triggered by either radiatively efficient accretion onto the black hole (the so-called ``cold mode''); or radiatively inefficient (``hot mode'') accretion \citet{BestHeckman12}. Powerful radio AGN are typically associated with the radiatively efficient mode, and often exhibit high-excitation ionization lines \citep{HardcastleEA07}. This mode of AGN triggering is often associated with mergers \citep{ShabalaEA12,RamosAlmeidaEA12}, and was likely prevalent in a younger, denser Universe. By contrast, the majority of low--redshift radio AGN are low power, and show no such strong emission lines; these are consistent with being fuelled by steady cooling of hot X--ray emitting gas in galactic haloes \citep{BestEA05,PopeEA12}.

From a theoretical viewpoint, AGN feedback is a crucial ingredient of all galaxy formation models,  truncating (the otherwise) excessive star formation in massive galaxies at late times, and ensuring that present-day ellipticals are ``red and dead'' \citep{CrotonEA06,BowerEA06}. This feedback can come from radiative pressure \citep{FabianEA06}, quasar winds \citep{NesvadbaEA08}, jets \citep{SA09b} or a combination of these \citep{HopkinsElvis10}.

A number of authors have argued that AGN feedback significantly affects the star formation histories of AGN hosts \citep{KavirajEA11,AntonuccioDeloguSilk08,KauffmannEA03}. \citet{RawlingsJarvis04} suggested on energetic grounds that a single AGN may be able to expel gas from multiple galactic haloes. Recently, \citet{SKS11} confirmed this observationally: shocks from the largest (hundreds of kpc--scale) radio AGN were found to truncate star formation in dwarf galaxies that themselves may have never hosted an AGN. This potentially important new mode of feedback depends sensitively on the combination of AGN jet power and environment density. In order to form a large radio AGN, the radio jet needs to pierce the dense environment of the AGN host without being disrupted by magnetohydrodynamic instabilities. Weak jets in moderate density environments, or moderate power jets in dense environments, are easily disrupted within dense galactic cores \citep{Alexander00}, and are thus unable to do large--scale feedback. Therefore, accurate measurement of AGN jet power is crucial to quantifying the effects of AGN on galaxy formation and evolution.

Measuring kinetic AGN power is difficult. Dynamical models, utilizing observed sizes and luminosities of radio AGN, can be used to estimate jet powers \citep{KDA97,SAAR08,AntogniniEA12}. These models rely on some knowledge of environment into which the radio lobes are expanding, and AGN age. A closely related cavity power method relies on measurements of the work done by the radio lobes in inflating cavities in the X-ray emitting gas \citep{RawlingsSaunders91,BirzanEA08}. This method also requires knowledge of both the density of the gas into which the lobes are expanding, and age of the radio lobes. Ages of synchrotron--emitting electrons can be estimated from either radio spectra \citep{AlexanderLeahy87} or dynamical models, and the two estimates typically differ by a factor of two \citep[e.g.][]{MachalskiEA04}. Furthermore, the cavity technique is limited to nearby, low-power radio galaxies, and the derived jet powers may thus be unrepresentative of the overall AGN population. Other sources of error in jet power estimates arise due to uncertainties in the exact gas density profile: for example, cavity measurements typically assume that the radio lobes expand into a constant density atmosphere throughout their lifetime.

In this paper, we apply a fundamentally different method for measuring AGN jet power to a sample of flat spectrum radio quasars. This method, first devised by \citet{Lobanov98} and further discussed by \citet{Hirotani05}, is based on the interpretation of the frequency dependent position shift of flat spectrum radio cores in terms of synchrotron self-absorption within the jet plasma. We compare the derived jet powers with narrow-line luminosities. 

Throughout the paper, we adopt the concordance cosmology of $\Omega_{\rm M}=0.27$, $\Omega_\Lambda=0.73$, $h=0.71$.

\section{Model and assumptions}
\label{sec:model}

This work is based on the standard model of self-absorbed radio cores, in which the unresolved, flat spectrum ``core" in VLBI images of radio quasars is understood to be associated with the part of the jet at which the optical depth is unity.  The core shift with frequency is interpreted as being due to the changing synchrotron self-absorption opacity in a freely expanding jet \citep{Marscher77,Konigl81}. Using this model, a measurement of the core shift between multiple frequencies along with a number of reasonable assumptions allows an estimate to be made of the physical parameters of the jet plasma, and therefore, an estimate of the jet kinetic energy flux. Here we recount the relevant details of the model, and expand on the analysis presented in \citet{Lobanov98} and \citet{Hirotani05} in the context of jet power measurements. 

\citet[][Appendix B]{SchwartzEA06} provides a detailed derivation of the equation for kinetic energy flux in a relativistic jet in terms of the physical parameters. The relevant quantities are defined as follows: $\beta=v/c$ is the jet bulk velocity and $\Gamma = (1-\beta^2)^{-1/2}$ the corresponding bulk Lorentz factor, $A$ the jet cross-sectional area, $n_e$ the density of relativistic electrons/positrons, $B$ the magnetic field strength, $u_e$ the energy density of relativistic electrons/positrons and  $u_B = B^2/8 \pi$ is the magnetic energy density in cgs units. We assume that the jet plasma may be a mixture of relativistically cold matter with pressure much less than rest mass energy density, and relativistic particles with pressure $p_e = \frac{1}{3} u_e$. The kinetic energy flux is then

\begin{equation}
  Q_{\rm jet} = \Gamma^2 c \beta A \left[ q_e u_e + q_B u_B + \frac{\Gamma-1}{\Gamma} q_r n_e m_e c^2 \right].
\label{eqn:Qjet_general}
\end{equation}
The parameters $q_e$ and $q_r$ define the relative contributions from cold matter to the internal energy density and rest-mass density, respectively. The parameter $q_e = \frac{4}{3}(1 + u_p/u_e)$ where $u_p$ is the internal energy density of cold matter (i.e. protons). The parameter $q_r = (1+\rho_p/\rho_e)$ where $\rho_p$ and $\rho_e$ are the rest mass density of cold matter, and relativistic electrons/positrons, respectively. In a purely electron/positron jet, $q_e = 4/3$ and $q_r = 1$. Finally, the parameter $q_B = 2 \frac{\langle B^2_\perp \rangle}{B^2}$ where $B_\perp$ is the component of magnetic field perpendicular to the jet velocity. $q_B$ depends on the magnetic field geometry, with $0 < q_B < 2$. Throughout this paper we assume the magnetic field is dominated by the perpendicular component, so that $q_B \approx 2$. 

\subsection{Geometry}
\label{sec:geometry}

We assume the quasar radio jet has a half opening angle $\phi$, and is viewed at an angle $\theta$ (typically $<10^\circ$) to the line of sight. The corresponding Doppler factor is $\delta=\left[ \Gamma \left(1- \beta \cos \theta  \right) \right]^{-1}$. The radial extent of the jet is described by $r$, while the transverse extent is given by $R = r \tan \phi \approx r \phi$.

\subsection{Magnetic field and particle distribution}
\label{sec:Bfield}

Following \citet{Konigl81} and \citet{Lobanov98} we parametrize the magnetic field as $B(r) = B_1 (r/{\rm 1 pc})^{-a}$; and the electron density as $n_e(r)=n_1 (r/{\rm 1 pc})^{-b}$; where $B_1$ and $n_1$ are the magnetic field strength and electron density 1 pc away from the central engine. Here, $r$ is the radial distance from the central engine. Observations suggest the core shift scales as the inverse of frequency ($1/\nu$; \citet{KovalevEA08,SokolovskyEA11}). This is indicative of conical jet structure and equipartition conditions, since parabolic jet structure produced due to external pressure gradients is likely to result in a significantly flatter core position--frequency dependence \citep{Lobanov98}. We therefore reasonably assume constant opening angle in our analysis. For such a jet, $1 < a < 2$, and the precise value depends on the magnetic field geometry: $a=2$ for the component of magnetic field parallel to the jet, and $a=1$ for the component of magnetic field perpendicular to the jet. Following \citet{Lobanov98} we assume that the perpendicular component of the magnetic field is the dominant one, such that $a=1$. Conservation of particles then implies $b = 2$. 

We assume a power-law energy distribution for the relativistic electrons/positrons of the form $N(\gamma)=N_0 \gamma^{-s}$. $N_0$ represents the scaling of the particle energy distribution, $s=1-2\alpha$ and $\alpha$ is the optically thin synchrotron spectral index, typically in the range $-0.5 < \alpha < -1$.

\subsection{Core shift}
\label{sec:coreShift}

For $a=1$, $b=2$, Hirotani (2005) gives the optical depth as:

\begin{eqnarray}
  \tau & = & 2.964\times 10^9 C(\alpha) n_1 B_1^{1.5-\alpha}\frac{-2\alpha}{\gamma_{\rm min}^{2\alpha}}\frac{\phi}{\sin\theta} \nonumber\\
  & & \times \left(\frac{1.759\times 10^7}{r\nu}\right)^{(2.5-\alpha)} \left(\frac{\delta}{2\pi(1+z)}\right)^{1.5-\alpha}
\label{eqn:Sm_tau}
\end{eqnarray}

Here, $\delta$ is the Doppler factor; $\nu$ is the observing frequency; $\alpha$ the spectral index; $n_1$, $B_1$ are electron densities and magnetic field strengths at 1 pc from the central engine; and all units are given in cgs except $r$ which is in parsecs. $C(\alpha)$ is a tabulated function of the spectral index. Setting the optical depth $\tau=1$ gives the location of the core as a function of frequency, $r(\nu)$:

\begin{eqnarray}
  r(\nu) & = & F(z,\alpha,\delta,\gamma_{\rm min},\phi,\theta) n_1^{\frac{1}{2.5-\alpha}} B_1^{\frac{1.5-\alpha}{2.5-\alpha}} \frac{1}{\nu} \nonumber \\
  F & = & 1.759\times 10^7 \left(2.964\times 10^9 C(\alpha)\frac{-2\alpha}{\gamma_{\rm min}^{2\alpha}}\frac{\phi}{\sin\theta}\right)^{\frac{1}{2.5-\alpha}} \nonumber\\
  & & \times \left(\frac{\delta}{2\pi(1+z)}\right)^{\frac{1.5-\alpha}{2.5-\alpha}}
\label{eqn:nu_s}
\end{eqnarray}

Equation~\ref{eqn:nu_s} can be rearranged to give the difference between radii (in parsecs) at which the core is observed at two frequencies,

\begin{equation}
  \Delta r_{\rm pc} = F n_1^{\frac{1}{2.5-\alpha}} B_1^{\frac{1.5-\alpha}{2.5-\alpha}} \left( \frac{\nu_1 - \nu_2}{\nu_2 \nu_1} \right)
  \label{eqn:Delta_r_pc}
\end{equation}

Equation \ref{eqn:Delta_r_pc} contains two unknowns, so in order to determine the physical parameters of the jet from a measured core shift, we must make some assumption about the relationship between particle density ($n_1$) and magnetic field strength ($B_1$) at one parsec, which we discuss in the following section.

\subsection{Calculating the magnetic field strength under the assumption of equipartition}
\label{sec:equipartition}

\citet{SokolovskyEA11} find that most sources for which core shifts are easily measured have $r(\nu) \propto \nu^{-1}$, consistent with equipartition of energy between particles and the magnetic field \citep{Lobanov98, SokolovskyEA11}. We thus reasonably assume that the magnetic field is in energy equipartition with the radiating particles ($u_B = u_e$). Then,
\begin{equation}
  n_1 = \frac{B_1^2}{8 \pi} \frac{1}{\langle \gamma \rangle m_e c^2} \nonumber
\label{eqn:n_0}
\end{equation}
where $\langle \gamma \rangle$ is the mean Lorentz factor of the relativistic electrons/positrons, and for our assumed power-law distribution
\begin{eqnarray}
\centering
\langle \gamma \rangle &=& \frac{\int_{\gamma_{\rm min}}^{\gamma_{\rm max}} \gamma N(\gamma) d\gamma}{ \int_{\gamma_{\rm min}}^{\gamma_{\rm max}}N(\gamma) d\gamma} \nonumber\\
& = & \begin{cases}
\centering
\gamma_{\rm min} \left( \frac{s-1}{s-2} \right) & \text{if } s > 2 \> \mbox{and} \> \gamma_{\rm max} \gg \gamma_{\rm min} \nonumber \\
\gamma_{\rm min} \ln \left( \frac{\gamma_{\rm max}}{\gamma_{\rm min}} \right) & \text{if } s = 2
\end{cases}
\end{eqnarray}
Hence, solving Equation~\ref{eqn:Delta_r_pc} for $B_1$ gives
\begin{equation}
B_1 = \left[  \frac{\Delta r_{\rm pc}}{F} \left(  \frac{\nu_1 \nu_2}{\nu_1 - \nu_2} \right) \right]^{\frac{2.5 - \alpha}{3.5 -\alpha}} \left[ \langle \gamma \rangle 8 \pi m_e c^2 \right]^{\frac{1}{3.5 - \alpha}}
\end{equation}

\subsection{Jet power}
\label{sec:jetPower}

Assuming a particular jet composition (electron--positron or electron--proton), spectral index $\alpha$, and Lorentz factor limits $\gamma_{\rm min}$ and $\gamma_{\rm max}$, from Equation~\ref{eqn:Qjet_general} we can write
\begin{eqnarray}
  Q_{\rm jet} & = & \kappa B_1^2 \Gamma^2 \beta \nonumber\\
  \kappa & = & 7.2 \times 10^{46} \phi^2  \left[ \frac{q_e}{q_B}  + 1 + \left( \frac{q_r}{q_B} \frac{\Gamma-1}{\Gamma} \frac{1}{\langle \gamma  \rangle} \right) \right] {\rm cm^3\,s^{-1}} \nonumber\\
  \label{eqn:Qjet_full}
\end{eqnarray}

Taking $\alpha = -0.5$, Equation~\ref{eqn:Qjet_full} yields
\begin{eqnarray}
Q_{\rm jet} & = & 2.4 \times 10^{26} \left[ \ln \left( \frac{\gamma_{\rm max}}{\gamma_{\rm min}} \right) \right]^{1/2} \frac{1}{(1+z)^2} \Theta\left( \Gamma, \theta, \phi \right)  \nonumber\\ && \times \left[\frac{q_e}{q_B}+1+ \left( \frac{q_r}{q_B} \frac{\Gamma-1}{\Gamma} \frac{1}{\langle \gamma \rangle} \right)  \right] \times \nonumber \\ &&  \times \left[\Delta r_{\rm mas} D_{\rm L}\left(\frac{\nu_1 \nu_2}{\nu_1-\nu_2}\right)\right]^{1.5} {\rm \,\,\, erg\,s^{-1}}
\label{eqn:Qjet}
\end{eqnarray}
where luminosity distance $D_{\rm L}$ is in Mpc, frequencies $\nu_1$ and $\nu_2$ in Hz, and core shift $\Delta r_{\rm mas}$ in milli-arcseconds. The constants
$q_e=4/3$, $q_r = 1$ and $q_B=2$ for an electron--positron jet in which the magnetic field is perpendicular to the jet, and $\Theta(\Gamma, \theta, \phi)=\frac{\Gamma^2 \beta}{\delta \sin \theta} \phi^{3/2}$ is a term corresponding to jet orientation, opening angle and velocity.

\section{Results and discussion}

\subsection{Correlation with narrow-line luminosity}

We use the above formalism to derive AGN jet powers for a sample of flat spectrum quasars with measured core shifts in the literature, and compare the derived jet powers to their narrow-line luminosities. In a seminal work, \citet{RawlingsSaunders91} found a correlation between narrow-line luminosity and jet power in a sample of 3CRR radio galaxies and steep spectrum quasars. For each source, these authors estimated jet power by calculating the work done on the surrounding gas by the radio lobes, and then dividing by the age of the radio source. This method assumes some coupling efficiency between the jet and the surrounding gas (factor $k$ in \citeauthor{RawlingsSaunders91}). Narrow-line emission results from the radiation field of the black hole accretion disk ionizing nearby gas clouds, and as such, a correlation between jet power and narrow-line luminosity is to be expected, since a higher accretion rate should result in both higher jet powers and stronger ionizing radiation field. 

From the literature, we selected quasars with measured core shifts, jet parameters and O[II] and O[III] line fluxes. Where multiple core shifts were available (as is the case for most of our sample), we calculated the frequency-independent core shift function $\Omega=\Delta r_{\rm mas} \left( \frac{\nu_1 \nu_2}{\nu_1 - \nu_2} \right)$ for each frequency pair, and adopted the median value. The jet power was then calculated using Equation~\ref{eqn:Qjet}. Narrow line luminosities were calculated via $L_{\rm NLR} = 3 (3 L_{\rm O [II]} + 1.5 L_{\rm O [III]})$ \citep{RawlingsSaunders91}. 

\begin{figure}
\begin{center}
\includegraphics[width=0.35\textwidth,angle=270]{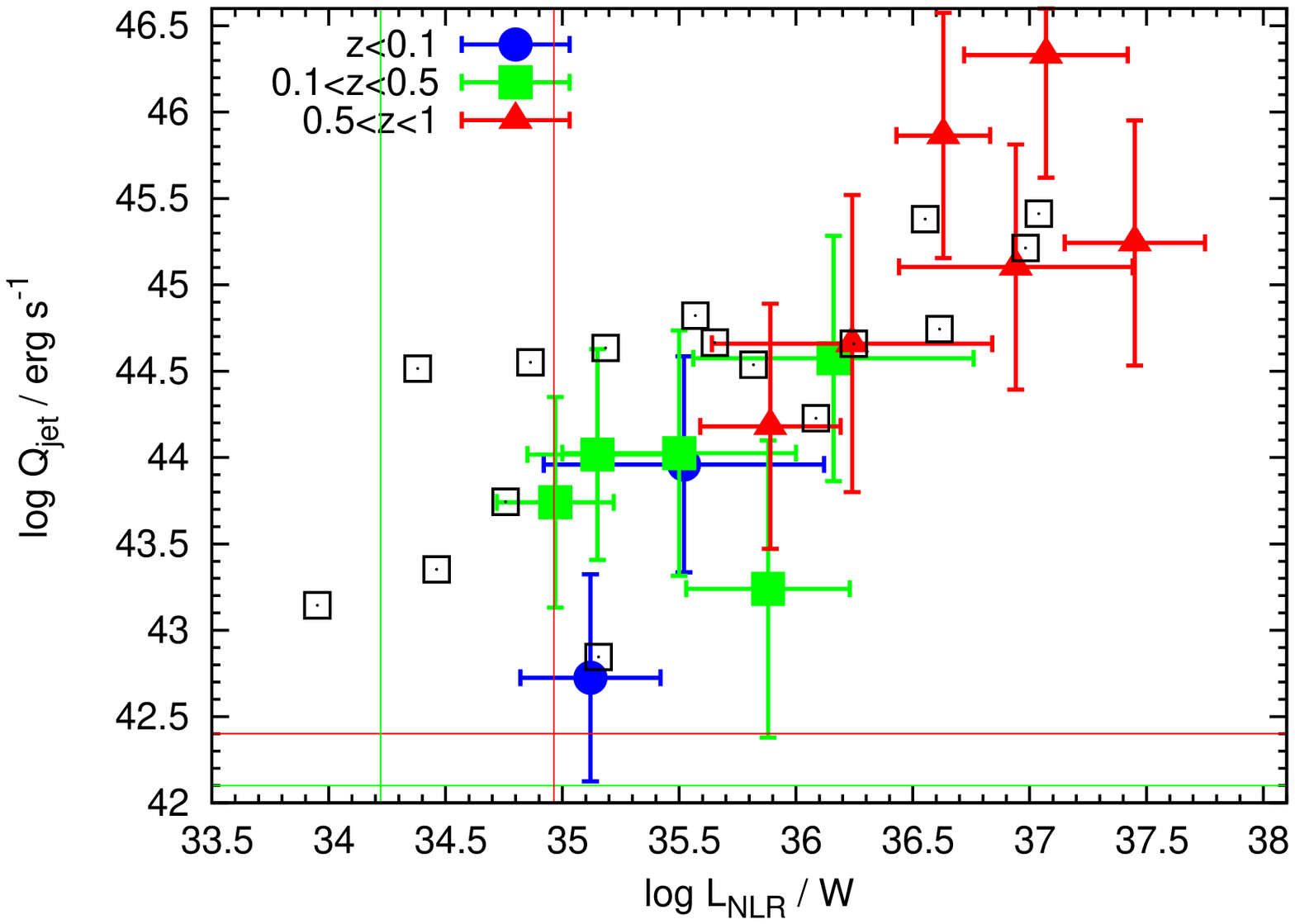}
\end{center}
\caption{Core shift jet power -- narrow-line luminosity relation (filled points) for a pair plasma. Core shifts are from: \citet{Lobanov98}, \citet{KovalevEA08}, \citet{SokolovskyEA11}, \citet{OSullivanGabuzda09}, \citet{PushkarevEA09}. Doppler factors and beaming angles are from: \citet{LahteenmakiValtaoja99}, \citet{XuEA99}, \citet{PushkarevEA09}. Line luminosities are from: \citet{LawrenceEA96}, \citet{WillottEA99}, \citet{GrimesEA04}. Open black squares show the results of \citet{RawlingsSaunders91} for a steep-spectrum sample of radio galaxies. The horizontal lines show conservative detection limits (0.1 mas between 2.3 and 8.4 GHz) in core shift at $z=0.5$ and $z=1$; and the vertical lines are similar limits in narrow-line luminosity. The detection limits at $z=0.1$ are too faint to appear on the plot. Jet powers and narrow-line luminosities in the upper right quadrant of the plot should be observable throughout our sample volume. Lack of data in the top left and bottom right sections of the plot suggest strongly that the correlation is real. Note that the normalization of both our and Rawlings \& Saunders relations depends on the assumed $\gamma_{\rm max}$ and $\gamma_{\rm min}$ cutoffs.}
\label{fig:jetPowers}
\end{figure}

Figure~\ref{fig:jetPowers} shows the resultant jet power -- narrow-line luminosity relation for an electron--positron jet. Also plotted are the \citet{RawlingsSaunders91} data converted to our cosmology. There is a strong correlation between the core shift--derived jet powers and narrow-line luminosity, and this correlation is consistent with the results of \citeauthor{RawlingsSaunders91}. Our slope of $1.0 \pm 0.2$ agrees with the Rawlings \& Saunders value of $0.9 \pm 0.2$. The scatter of 0.6 dex about the best-fit relation is also comparable with the Rawlings \& Saunders value of 0.5 dex. It is worth emphasizing that those authors used a completely different method (cavity jet powers) and a completely different sample (steep spectrum) of radio sources.

\subsection{Comparison with other jet power estimates}

A number of authors have used alternative techniques to estimate jet power. \citet{CavagnoloEA10} built on the work of \citet{BirzanEA08} in reporting a correlation between cavity-derived AGN jet powers and radio luminosities. \citet{WillottEA99} similarly use AGN radio luminosity as a proxy for jet power, based on a correlation between narrow-line and radio luminosities. These methods are useful, largely because radio luminosity is a relatively easy quantity to measure; and yield similar scatter (0.8 and 0.5 dex respectively) to our core shift technique. However, one must be careful in extrapolating these results to large samples. Dynamical models of radio sources, on which the \citet{WillottEA99} method is based, only apply to powerful Fanaroff--Riley (FR) type II radio sources. On the other hand, weaker FR I radio sources dominate AGN counts in the local Universe \citep{BestEA05,SAAR08}. Furthermore, these models are sensitive to environment: a jet with a given kinetic power will appear brighter at radio wavelengths if it propagates though denser gas. By contrast, the core shift method is independent of environment.

The slope of the cavity jet power -- integrated radio power relation reported by \citet{CavagnoloEA10} is $0.75 \pm 0.14$ at 1.4 GHz, and $0.64 \pm 0.09$ at 327 MHz. At low frequencies the AGN radio luminosity will be dominated by lobe emission; while at higher frequencies flat-spectrum components such as jet core and hotspots will also contribute. This flat-spectrum contribution will be greater for powerful jets, resulting in a flattening of the $Q_{\rm jet}$--$L_{\rm radio}$ slope at higher frequencies. These factors, along with environmental dependence, are also likely to be responsible for the larger scatter (0.8 dex) in the \citet{CavagnoloEA10} relation compared with our method. Nevertheless, the slopes reported by these authors are broadly consistent with our results.

\subsection{Estimating core shift jet power}

\begin{figure}
\begin{center}
\includegraphics[width=0.35\textwidth,angle=270]{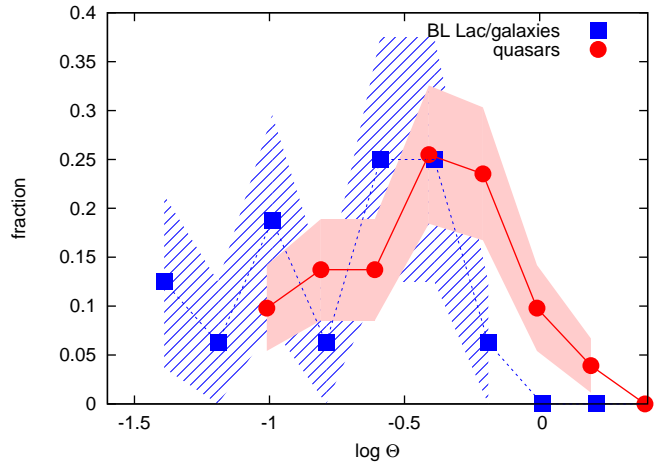}
\end{center}
\caption{Distribution of beaming parameter $\Theta$ for the complete sample of Pushkarev et al. (2009). Red circles represent quasars (51 objects), blue squares are BL Lacs and galaxies (16 objects). Shaded regions show Poisson uncertainties.}
\label{fig:PhiDist}
\end{figure}

\citet{PushkarevEA09} derived jet properties for a large sample (67) of flat spectrum radio AGN. This sample is complete to 1.5 Jy at 15 GHz \citep{ListerEA09}, and is thus representative of the flat spectrum radio AGN population. Using this data, we find that the distribution of the function $\Theta(\Gamma, \theta, \phi)$ is roughly log-normal for quasars (Figure~\ref{fig:PhiDist}), which constitute the majority of the sample. Quasars have a mean of $\left< \log \Theta \right> = -0.55$ and standard deviation of $\sigma_{\log \Theta} = 0.37$. Due to insufficient statistics (16 objects) it is difficult to say whether the distribution of $\Theta$ for BL Lacs/galaxies is similarly log-normal. Values of $\left< \log \Theta \right> = -0.85$ and $\sigma_{\log \Theta} = 0.42$ are obtained under this assumption.

\begin{figure}
\begin{center}
\includegraphics[width=0.35\textwidth,angle=270]{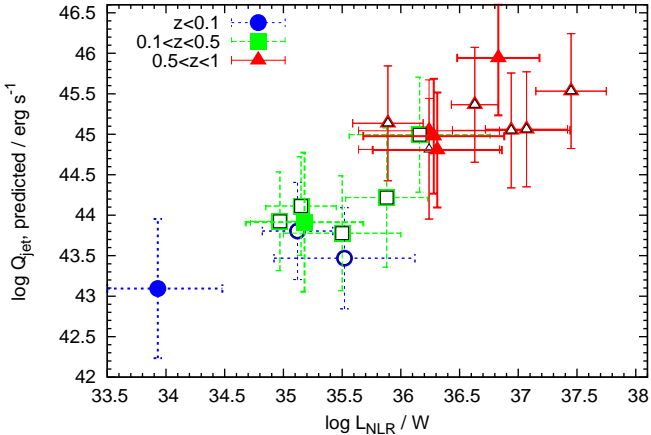}
\end{center}
\caption{Jet powers predicted using Equation~\ref{eqn:QjetAvg}. Open symbols represent AGN with known jet parameters, plotted in Figure~\ref{fig:jetPowers}. Filled symbols are AGN for which beaming parameters are not known. Core shifts can be used to estimate jet power even if the jet parameters are unavailable.}
\label{fig:coreShiftFunction}
\end{figure}

Importantly, these distributions are narrow enough for core shift measurements to be useful predictors of jet power even if jet orientation parameters are not known: for a given narrow-line luminosity, 68 percent of the AGN should have jet powers that are within a factor of 2.5 from the ``best guess'' values. This is illustrated in Figure~\ref{fig:coreShiftFunction}, where we plot the predicted jet power $Q_{\rm jet, \,approx}$, which is calculated using the mean $\log \Theta$ values above, against narrow-line luminosity. Unlike Figure~\ref{fig:jetPowers}, AGN with unknown values of $\Gamma$, $\theta$ and $\phi$ are also plotted. The strong correlation between these two quantities (slope $0.8 \pm 0.1$, scatter 0.4 dex) suggests that core shifts can be used to estimate the kinetic energy flux of flat spectrum AGN even if the jet velocity, viewing angle, and opening angle are not known. The best--fit relation for quasars, assuming an electron--positron jet in equipartition for $\alpha=-0.5$, is

\begin{eqnarray}
Q_{\rm e^+e^-,\ approx} & = & 6.8^{+9.2}_{-3.8} \times 10^{25} \left[ \ln \left(\frac{\gamma_{\rm max}}{\gamma_{\rm min}}\right) \right]^{1/2} \frac{1}{(1+z)^2} \nonumber\\ && \times \left[\Delta r_{\rm mas} D_{\rm L} \left(\frac{\nu_1 \nu_2}{\nu_1-\nu_2}\right)\right]^{3/2} {\rm \,\,\,erg\,s^{-1}}
\label{eqn:QjetAvg}
\end{eqnarray}
where again luminosity distance $D_{\rm L}$ is in Mpc, frequencies $\nu_1$ and $\nu_2$ in Hz, and core shift $\Delta r_{\rm mas}$ in milli-arcseconds. The median jet power is a factor of 2 lower for BL Lacs, for the same core shift and other properties, due to the different distribution of jet parameters (as given by $\Theta$; see Figure~\ref{fig:PhiDist}).

For an electron--proton jet, the rest mass term in Equation~\ref{eqn:Qjet} is important. For normal cosmic plasma, $q_r \approx 2200$ \citep{SchwartzEA06}, and thus
\begin{equation}
Q_{\rm p^+e^-,\ approx} = \left[ 1+\frac{660}{\gamma_{\rm min} \ln{ \left( \frac{\gamma_{\rm max}}{\gamma_{\rm min}} \right)} } \right] Q_{\rm e^+e^-,\ approx}
\label{eqn:QjetProton}
\end{equation}
The minimum Lorentz factor $\gamma_{\rm min}$ in parsec-scale jets is highly uncertain, and whilst the minimum Lorentz factor in the hotspots of radio galaxies is up to $10^3$ in some sources, $\gamma_{\rm min}$ in the jets prior to reaching the termination shock could be much lower \citep{GodfreyEA09}. For $\gamma_{\rm min}=10$ and $\frac{\gamma_{\rm max}}{\gamma_{\rm min}}=100$ the estimated electron--proton jet power is a factor of 15 greater than the electron--positron case. This method could potentially be used to place constraints on the jet composition, and the value of $\gamma_{\rm min}$ in parsec-scale jets. 

The estimation of jet powers from limited data, as given by Equations~\ref{eqn:QjetAvg} and \ref{eqn:QjetProton}, is possible because the flat spectrum constraint necessarily limits the AGN sample to objects with a narrow range of viewing angle, and $\Gamma$ and $\phi$ have narrow distributions.

\section{Conclusions}

We have calculated AGN jet powers from multi-frequency core shifts in flat spectrum quasars. We find a strong correlation between jet power and AGN narrow-line luminosity, consistent with the results of \citet{RawlingsSaunders91} for steep spectrum radio galaxies. The core shift method has strong predictive power, allowing accurate estimates of jet power to be made in the absence of detailed jet and AGN orientation parameters.

The geometry of the Universe makes the core shift method a potentially very powerful one for measuring AGN jet powers, particularly in the era of new, high sensitivity instruments capable of very high astrometric precision such as the Square Kilometre Array \citep{GodfreyEA12}. The angular diameter distance turns over at redshift 1.6, making it in principle even easier to measure a core shift at redshift 3 than at redshift 1.5. This is also true for other jet power measurement methods, such as dynamical and cavity models. However, the {\it luminosity} distance increases with redshift, making the detection of steep spectrum objects difficult. The advantage of the core shift method is in the fact that the flat spectrum nature of core--shifting quasars allows Doppler boosting to bring even intrinsically faint radio sources above the detection limit. Thus, the core shift method will be able to probe higher redshifts and lower luminosities than traditional jet power measurement methods. In addition to measuring AGN jet power, this technique is potentially useful in studying the assembly and evolution of black holes in the very early Universe, and constraining the composition and the value of $\gamma_{\rm min}$ in parsec-scale jets.

\acknowledgements{}
We thank the anonymous referee for useful suggestions. SS thanks the ARC for a Super Science Fellowship. JS is grateful to the University of Tasmania for a Dean's Summer Research Scholarship. LG thanks Curtin University for a research fellowship. This research has made use of Ned Wright's cosmology calculator \citep{Wright06}; and of the NASA/IPAC Extragalactic Database (NED) which is operated by the Jet Propulsion Laboratory, California Institute of Technology, under contract with the National Aeronautics and Space Administration.

\end{document}